\documentstyle[12pt] {article}
 
\begin{document}
 \title{ {\small { \bf RADIATIVE $B^{*}\rightarrow B\gamma$ and
 $D^{*}\rightarrow D\gamma$ DECAYS IN LIGHT CONE QCD SUM RULES.}}}
 \author{ {\small T.M.ALIEV \,\,, D. A. DEMIR \,\,, E.ILTAN \,\,,and N.K.PAK}
\\ {\it {\small Physics Department, Middle East Technical University}}
\\ {\it {\small Ankara,Turkey}}}
 \begin{titlepage}
 \maketitle
 \thispagestyle{empty}
 \begin{abstract}
 \baselineskip  .7cm
 The radiative decays $ B^{*} (D^{*})\rightarrow B(D) \gamma$ are
 investigated in the framework of light cone QCD sum rules.
 The transition amplitude and decay rates are estimated.It is shown that
 our results on branching ratios of D meson decays are in good agreement
 with the existing experimental data.
 \end{abstract}
 \end{titlepage}
 \baselineskip  .7cm
  \newpage
 \section{Introduction}
One of the main goals of the future B meson and charm- $\tau$ factories
is a deeper and
more comprehensive investigation of the B- and D-meson physics.

The radiative  decays constitute an important for a
comprehensive study the properties of the new meson states containing
heavy quark. However, for interpretation of the experimental results we
immediately deal with large distance effects. It is well known that QCD
sum rules method [1] take into account such large distance effects and
a powerfull tool for investigating the properties of hadrons.
Nowadays, in current literature,  an alternative to
"classical QCD sum rules method", the QCD sum rules based on light cone
expansion is widely exploited. Main features of this version is that it
is based on the
approximate conformal non-perturbative invariance of QCD, and
instead of many process-dependent non-perturbative parameters
in classical "QCD sum rules", it involves a new universal non-perturbative
parameter, namely the wave function [3].
This sum rules previously were succesful applied for calculation
the decay amplitude $\Sigma\rightarrow p\gamma$ [4], the nucleon magnetic
moment, $g_{\pi NN}$ and $g_{\rho\omega\pi}$ couplings [5], form
factors of semileptonic and radiative decays [6-9],
the $\pi A\gamma^{*}$ form factor
[10], $B\rightarrow \rho\gamma$ and $D\rightarrow \rho\gamma$
decays [11,12], $B^{*}B\pi$ and $D^{*}D\pi$ coupling constants [13] etc.
In this work we study the radiative $B^{*}(D^{*})\rightarrow B(D) \gamma$
decays in the framework
of the light cone QCD sum rules.
Note that these decays were investigated
Ref [14,15], in the framework of classical QCD sum rules method.

The paper is organized as follows.In section 2, we derive the sum rules
which describes $B^{*}(D^{*})\rightarrow B(D)\gamma$ in the framework
of the light-cone sum rules. In the last section we present the numerical
analysis.

\section{The Radiative $B^{*}\rightarrow B\gamma$ decay}
According to the general strategy of QCD sum rules, we want to obtain
the transition amplitude for $B^{*}\rightarrow B\gamma$ decay,
by equating the representation
of a suitable correlator function in hadronic and quark-gluon language.
For this aim we consider the following correlator
\begin{eqnarray}
\Pi_{\mu}(p,q) & = & i\int d^{4}x e^{ipx}
<0|T[\bar{q}(x)\gamma_{\mu} b(x),\bar{b}(0)i\gamma_{5}q(0)]|0>_{F}
\end{eqnarray}
in the external electromagnetic field
\begin{eqnarray}
F_{\alpha\beta}(x)=i(\epsilon_{\beta}q_{\alpha}-\epsilon_{\alpha}q_{\beta})
e^{iqx}
\end{eqnarray}
Here q is the momentum and $\epsilon_{\mu}$ is the polarization vector
of the electromagnetic field.
The Lorentz decomposition of the correlator is
\begin{eqnarray}
\Pi_{\mu}=i\epsilon_{\mu\nu\alpha\beta}p_{\nu}\epsilon_{\alpha}q_{\beta} \Pi
\end{eqnarray}
Our main problem is to calculate $\Pi$ in Eq. (3). This problem can be solved
in the Euclidean space where both, $p^{2}$ and
$p'^{2}=(p+q)^{2}$ are negative and large.
In this deep Euclidean region, photon interact with the
heavy quark perturbatively. The various contributions to the correlator
function Eq.(1) are depicted in Fiq. (1), where diagrams (a) and (b)
represent
the perturbative contributions, (c) is the quark condensate, (d) is the
5-dimensional operator, (e) is the photon interaction with soft quark
line (i.e. large distance
effects), and (f) is the three particle high twist contributions.
A part of calculations of these diagrams  was performed in [14,15].

First, let us calculate the perturbative contributions, namely diagram
(b). After standard calculation of the bare loop we have
 \begin{eqnarray}
 \Pi_{1} &=& \frac{Q_{q}}{4 \pi^{2}}N_{c}\int_{0}^{1} x dx
    \int_{0}^{1} dy \frac{m_{b}\bar{x}+m_{a}x}{m_{a}^{2}x+m_{b}^{2}
    \bar{x}-p^{2}x\bar{x}y- p'^{2}x\bar{x}\bar{y}}
%    \\
%    A_{2}^{a}&=&A_{1}(m_{c} \leftrightarrow m_{a}, Q_{a}\leftrightarrow
%    Q_{c})
    \end{eqnarray}
    where $N_{c}=3$ is the color factor, $\bar{x}=1-x$ ,$\bar{y}=1-y$,
$p'=p+q$, $Q_{q}$ and $m_{a}$ are the charge and the mass of the light quarks.
Note that the contribution of the diagram (a) can be obtained
by making the following replacements in Eq. (4) :
$ m_{b} \leftrightarrow m_{a}, e_{q}\leftrightarrow  e_{Q}$.
The next step is to use the exponetial representation for the denominator
\begin{eqnarray}
\frac{1}{C^{n}}=\frac{1}{(n-1)!}\int_{0}^{\infty} d\alpha \alpha^{n-1}
e^{-\alpha C} \nonumber
\end{eqnarray}
Then
\begin{eqnarray}
\Pi_{1}=\frac{Q_{q}N_{c}}{4\pi^{2}} \int_{0}^{1} x dx \int_{0}^{1} dy
[m_{b}\bar{x}+m_{a}x] \int_{0}^{\infty} d\alpha e^{(
m_{a}^{2}x+m_{b}^{2}\bar{x}-p^{2}x\bar{x}y-p'^{2}x\bar{x}\bar{y})}
\end{eqnarray}
Application of the double Borel operator $\hat B(M_{1}^{2})
\hat B(M_{2}^{2})$ on $\Pi_{1}$ gives
\begin{equation}
\tilde{\Pi}_{1} = \frac{Q_{q}N_{c}}{4\pi^{2}}
\frac{\sigma_{1}\sigma_{2}}{\sigma_{1}+\sigma_{2}}\int_{0}^{1} dx
\frac{1}{\bar{x}}
[m_{b}\bar{x}+m_{a}x] \, exp \{- \frac{m_{a}^{2}x+m_{b}^{2}\bar{x}}
{x\bar{x}}
(\sigma_{1}+\sigma_{2}) \}
\end{equation}
where $\sigma_{1}= \frac{1}{M_{1}^{2}}$ and $\sigma_{2}=\frac{1}{M_{2}^{2}}$.
In deriving Eq. (6) we use the definition
\begin{eqnarray}
\hat{B}(M^{2})e^{-\alpha^{2}}=\delta (1-\alpha M^{2})
\end{eqnarray}
Now consider the combination
\begin{equation}
\frac{1}{st} \hat{B}(\frac{1}{s},\sigma_{1})\hat{B}(\frac{1}{s},
\sigma_{2})\frac{\tilde {\Pi}_{1}}{\sigma_{1}\sigma_{2}}
\end{equation}
which just gives the spectral density [16].
Using Eq. (6) and Eq. (8), for the spectral density, we get
\begin{eqnarray}
\rho_{1}(s,t)&=& \frac{Q_{q}N_{c}}{4\pi^{2}}\int_{x_{0}}^{x_{1}} dx
\delta(s-t)\theta(s-(m_{b}+m_{a})^{2})
\theta(t-(m_{b}+m_{a})^{2}) \nonumber \\ &.&\frac{(m_{b}\bar{x}+m_{a}x)}
{\bar{x}}
\end{eqnarray}
where the integration region is determined by the inequality
\begin{equation}
sx \bar{x}-(m_{b}^{2}\bar{x}+m_{a}^{2}x) \geq 0
\end{equation}
Using Eq. (9) for the spectral density, we get
\begin{eqnarray}
\rho_{1}^{a}(s,t)&=& \frac{Q_{q}N_{c}}{4\pi^{2}}\delta(s-t)\theta
(s-(m_{b}+m_{a})^{2})
\theta (t-(m_{b}+m_{a})^{2}) \nonumber \\ \{ &-&(m_{b}-m_{a})
\lambda (1,\kappa,l)+m_{b}ln\frac{1+\kappa-l+\lambda(1,\kappa,l)}
{1+\kappa-l-\lambda(1,\kappa,l)} \} \\
\rho_{2}(s,t)&=&\rho_{1}(a\leftrightarrow
b,m_{a}\leftrightarrow m_{b},Q_{a}\leftrightarrow Q_{b})
\end{eqnarray}
where $ \kappa=\frac{m_{a}^{2}}{s}, l=\frac{m_{b}^{2}}{s}$ and
\begin{equation}
\lambda(1,\kappa,l)=\sqrt{1+\kappa^{2}+l^{2}-2\kappa-2l-2\kappa l}
\end{equation}
Finally for the perturbative part of the correlator we have
\begin{eqnarray}
\Pi^{Per}=\frac{N_{c}}{4\pi^{2}}\int ds \frac{1}{(s-p^{2})(s-p'^{2})}
[(Q_{a}-Q_{b})(1-\frac{m_{b}^{2}}{s})+Q_{b}ln\frac{s}{m_{b}^{2}}]
\end{eqnarray}
Here we have neglected the mass of the light-quark. Applying Eq.
(7), the Borel transformation, we get
\begin{eqnarray}
\Pi^{Per}=\frac{N_{c}}{M_{1}^{2}M_{2}^{2}4\pi^{2}}\int ds
e^{-s(\frac{1}{M_{2}^{2}}+\frac{1}{M_{2}^{2}})}
[(Q_{a}-Q_{b})(1-\frac{m_{b}^{2}}{s})+Q_{b}ln\frac{s}{m_{b}^{2}}]
\end{eqnarray}

After simple calculation, for the double Borel transformed quark condensate
contribution, we have:
\begin{eqnarray}
\Pi^{\bar{q}q}=Q_{b}<\bar{q}q>\frac{1}{M_{1}^{2}M_{2}^{2}}
e^{-\frac{m_{b}^{2}}{M_{1}^{2}+M_{2}^{2}}}
\end{eqnarray}
and the 5-dimensional operator contribution is
\begin{eqnarray}
\Pi^{\bar{q}q}&=& Q_{b}<\bar{q}q>\frac{1}{M_{1}^{2}M_{2}^{2}}
e^{-\frac{m_{b}^{2}}{M_{1}^{2}+M_{2}^{2}}} \nonumber \\
&.& \{ -\frac{m_{0}^{2}m_{b}^{2}}{4}( \frac{1}{M_{1}^{2}}+\frac{1}
{M_{2}^{2}})^{2}+\frac{1}{3}\frac{m_{0}^{2}}{M_{2}^{2}} \}
\end{eqnarray}

For the calculation of the diagram corresponding to the propogation
of the soft quark in the external electromagnetic
field, we use the light cone expansion for non-local operators.
After contracting the b-quark line in Eq(1) we get
\begin{eqnarray}
\Pi_{\mu}=i\int d^{4}x \frac{d^{4}k}{(2\pi)^{4}i}
\frac{e^{i(p-k)x}}{m_{b}^{2}-k^{2}}
<0|\bar{q}(x)\gamma_{\mu}(m_{b}+\not\!k)\gamma_{5}q(x)|0>_{F}
\end{eqnarray}
Using the identity $\gamma_{\mu}\gamma_{\alpha}\gamma_{5}=g_{\mu\alpha}
\gamma_{5}+i\sigma_{\rho\beta}\epsilon_{\mu\alpha\rho\beta}$
Eq. (18) can be written as
\begin{eqnarray}
\Pi_{\mu}=-\epsilon_{\mu\alpha\rho\beta}\int d^{4}x \frac{d^{4}k}
{(2\pi)^{4}i}\frac{e^{i(p-k)x}k_{\alpha}}{m_{b}^{2}-k^{2}}
<0|\bar{q}(x)\sigma_{\rho\beta}q|0>_{F}
\end{eqnarray}
The leading twist-two contribution to this matrix element in the
presence of the external electromagnetic field is defined as [4]:
\begin{eqnarray}
<\bar{q}(x)\sigma_{\rho\beta}q>_{F}=Q_{q}<\bar{q}q>\int_{0}^{1}du \phi (u)
F_{\alpha\beta}(ux)
\end{eqnarray}
Here the function $\phi (u)$ is the photon wave function. The asymptotic
form of this wave function is well known [4,17,18] to be,
\begin{eqnarray}
\phi (u)=6\xi u(1-u)
\end{eqnarray}
where $\xi$ is the magnetic susceptibility.

The most general decomposition of the relevant matrix element, to the
twist-4 accuracy, involves two new invariant functions
(see for example [11,12]):
\begin{eqnarray}
<\bar{q}(x)\sigma_{\rho\beta}q>_{F}&=&Q_{q}<\bar{q}q> \{ \int_{0}^{1}du
x^{2}\phi_{1} (u)F_{\rho\beta}(ux) \nonumber \\ &+&
\int_{0}^{1}du
\phi_{2} (u) [ x_{\beta}x_{\eta}F_{\rho\eta}(ux) \nonumber \\
&-& x_{\rho}x_{\eta}
F_{\beta\eta}(ux)-x^{2}F_{\rho\beta}(ux)]  \}
\end{eqnarray}

In [11] it was shown that
\begin{eqnarray}
\phi_{1}(u)&=& -\frac{1}{8} (1-u)(3-u) \nonumber \\
\phi_{2}(u)&=& -\frac{1}{4}  (1-u)^{2}
\end{eqnarray}
So, for twist 2 and 4 contributions we get
\begin{eqnarray}
\Pi^{twist 2 +twist 4}&=&Q_{q}<\bar{q}q> \{ \int _{0}^{1}\frac{\phi (u)
du}{m_{b}^{2}-
(p+uq)^{2}}\nonumber \\&-&
4\int _{0}^{1}\frac{(\phi_{1}(u)-\phi_{2}(u)) du}{(m_{b}^{2}-
(p+uq)^{2})^{2}}[1+\frac{2m_{b}^{2}}{(m_{b}^{2}-(p+uq)^{2})^{2}}] \}
\end{eqnarray}
In order to perform the double Borel transformation we rewrite denominator
in the following way:
\begin{eqnarray}
m_{b}^{2}-(p+uq)^{2}=m_{b}^{2}-(1-u)p^{2}-(p+q)^{2}u \nonumber
\end{eqnarray}
and applying the Wick rotation
\begin{eqnarray}
m_{b}^{2}-(p+uq)^{2}\rightarrow m_{b}^{2}+(1-u)p^{2}+(p+q)^{2}u
\nonumber
\end{eqnarray}
Using the exponential representation for the denominator we get
\begin{eqnarray}
\Pi^{twist 2 +twist 4}&=& e_{q}< \bar{q}q>e^{-m_{b}^{2}(\frac{1}{M_{1}^{2}}+
\frac{1}{M_{2}^{2}})} [\phi (\frac{M_{1}^{2}}{M_{1}^{2}+M_{2}^{2}})
\frac{1}{M_{1}^{2}+M_{2}^{2}}\nonumber \\&-&
4(\phi_{1}(\frac{M_{1}^{2}}{M_{1}^{2}+M_{2}^{2}})
-\phi_{2}(\frac{M_{1}^{2}}{M_{1}^{2}+M_{2}^{2}}) )
\nonumber \\ &.& (\frac{1}{M_{1}^{2}M_{2}^{2}}+
\frac{m_{b}^{2} (M_{1}^{2}+M_{2}^{2})}{M_{1}^{4}M_{2}^{4}})]
\end{eqnarray}
The mass of $B^{*}(D^{*})$ and $B(D)$ mesons are practically equal.
So, it is natural to take $M_{1}^{2}=M_{2}^{2}$, and introduce new Borel
parameter $M^{2}$ such that $M_{1}^{2}=M_{2}^{2}=2M^{2}$.
In this case the theoretical part of the sum rules become
\begin{eqnarray}
\Pi^{theor}&=&\frac{3}{4\pi^{2}}\int_{m_{b}^{2}}^{s_{0}} ds
e^{-s\frac{1}{M^{2}}}
[(Q_{q}-Q_{b})(1-\frac{m_{b}^{2}}{s})+Q_{b}ln\frac{s}{m_{b}^{2}}]
\frac{1}{4M^{4}} \nonumber\\
&+& Q_{b}<\bar{q}q>e^{-\frac{m_{b}^{2}}{M^{2}}}[1-
\frac{m_{0}^{2}m_{b}^{2}}{M^{4}}+\frac{m_{0}^{2}}{6M^{2}}]
\frac{1}{4M^{4}} \nonumber\\
&+&  Q_{q}< \bar{q}q> (e^{-m_{b}^{2}/M^{2}}-e^{-s_{0}/M^{2}})
[M^{2}\phi (\frac{1}{2}) \nonumber \\
&-& 4(1+\frac{m_{b}^{2}}{M^{2}})
(\phi_{1}(1/2)-\phi_{2}(1/2) ] \frac{1}{4M^{4}}
\end{eqnarray}
In the derivation of Eq. (26), we have subtracted the continuum and higher
resonance states from the double  spectral density. The details of this
procedure are given in [13].

For constructing the sum rules we need the expression for the physical
part as well. Saturating Eq.(1) by the lowest lying meson states, we have
\begin{eqnarray}
\Pi_{\mu}^{phys}= \frac{<0|\bar{q}\gamma_{\mu}b|B^{*}><B^{*}|B\gamma>
<B|\bar{b}i\gamma_{5}q|0>}{(m^{2}_{B^{*}}-p^{2})(m^{2}_{B}-(p+q)^{2})}
\end{eqnarray}
These matrix elements are defined as
\begin{eqnarray}
<0|\bar{q}\gamma_{\mu}b|B^{*}> &=& \epsilon_{\mu} f_{B^{*}}m_{B^{*}} \\
<B|\bar{b}i\gamma_{5}q|0>&=& \frac{f_{B}m^{2}_{B}}{m_{b}} \\
<B^{*}|B\gamma>&=& \varepsilon_{\alpha\beta\rho\sigma}
p_{\alpha}q_{\beta}\epsilon_{\rho}^{*} \epsilon^{*(\gamma)}_{\rho}h/m_{B}
\end{eqnarray}
Here $\it h$ is the dimensionless amplitude for the transition matrix
element,
$\epsilon_{\mu}$ , and $m_{B^{*}} $ are the polarization four-vector and
the mass of the vector particle respectively, $f_{B}$ is the leptonic
decay constant and $ m_{B}$ is the mass of the pseudoscalar particle,
$q_{\beta}$ and $\epsilon_{\sigma}$ are the photon momentum and the
polarization vector.
Applying the double Borel transformation we get for the physical part
of the sum rules
\begin{eqnarray}
\Pi^{phys}=f_{B^{*}}m_{B^{*}}f_{B}m_{B}\frac{h}{m_{b}}
\frac{e^{-(m^{2}_{B^{*}}+m^{2}_{B})/2M^{2}}}{4M^{4}}
\end{eqnarray}
Note that the contribution of three-particle twist-4 operators are very
small [4], and thus we neglect them (Fig. (1)).
{}From Eqs.(26-30) we get the dimensionless coupling constant as
\begin{eqnarray}
h&=&\frac{m_{b}}{f_{B^{*}}m_{B^{*}}f_{B}m_{B}}
e^{(m^{2}_{B^{*}}+m^{2}_{B})/2M^{2}}\nonumber \\
&.& \{ \frac{3}{4\pi^{2}}\int_{m_{b}^{2}}^{s_{0}} ds
e^{-s\frac{1}{M^{2}}}
[(Q_{q}-Q_{b})(1-\frac{m_{b}^{2}}{s})+Q_{b}ln\frac{s}{m_{b}^{2}}]\nonumber\\
&+& <\bar{q}q>e^{-\frac{m_{b}^{2}}{M^{2}}}[Q_{b} (1-
\frac{m_{0}^{2}m_{b}^{2}}{8M^{4}}+\frac{m_{0}^{2}}{6M^{2}}]\nonumber\\
&+& (e^{-m_{b}^{2}/M^{2}}-e^{-s_{0}/M^{2}})
[Q_{q}\phi (\frac{1}{2})M^{2}\nonumber \\
&-&4Q_{q}(1+\frac{m_{b}^{2}}{M^{2}})
(\phi_{1}(1/2)-\phi_{2}(1/2))] \}
\end{eqnarray}
\section{Numerical Analysis of the Sum rules}
The main issue concerning of Eq. (32) is the determination of the
dimensionless transition amplitude $ \it h $.
First,we give a summary of the parameters entering in the sum rules
Eq. (32).
The value of the magnetic susceptibility of the medium in the presence
of external field was determined in [19,20]
\begin{eqnarray}
\chi(\mu^{2}=1GeV^{2}) =-4.4GeV^{-2}\nonumber
\end{eqnarray}
If we include the anomalous dimension  of the current
$\bar{q}\sigma_{\alpha\beta}q$, which is $(-4/27)$ at $\mu=m_{b}$ scale,
we get
\begin{eqnarray}
\chi (\mu^{2}=m_{b}^{2}=-3.4GeV^{-2}\nonumber
\end{eqnarray}
and
\begin{eqnarray}
<\bar{q}q>=-(0.26\, GeV)^{3}\nonumber
\end{eqnarray}
The leptonic decay constants $f_{B(D)}$ and  $f_{B^{*}(D^{*})}$
are known from two-point sum rules related to $B(D)$ meson channels:
$f_{B(D)}=0.14\, (0.17)\,\, MeV$ [13,21] , $f_{B^{*}(D^{*})}=0.16\, (0.24)
\,\,GeV$  [13,22,23,24], $m_{b}=4.7 \,\,GeV$ ,
$m_{u}=m_{d}=0$, $m_{0}^{2}=(0.8\pm0.2)\,\,GeV^{2} $,
$m_{B^{*}(D^{*})}=5.\,\, (2.007)\,\, GeV$ and
$m_{B(D)}=5.\,\, (1.864)\,\, GeV \,$, and
for continuum threshold we choose $s_{0}^{B}(s_{0}^{D}) =36\,(6)\,\,GeV^{2}$.

The value  $\phi_{1}(u)-\phi_{2}(u)$ is calculated in [11]:
\begin{eqnarray}
\phi_{1}(u)-\phi_{2}(u)=\frac{-1}{8}(1-u^{2})\nonumber
\end{eqnarray}
{}From the asymptotic form of the photon wave function, given in Eq. (21),
we get
\begin{eqnarray}
\phi(1/2)=3/2\chi
\end{eqnarray}

Having fixed the input parameters,
it is necessary to find a range of $M^{2}$  for which the sum
rule is reliable.The lowest value of $M^{2}$, according to
QCD sum rules ideology, is determined by
requirement that the power corrections are reasonably small.
The upper bound is determined by the condition that the continuum
and the higher states contributions remain under control.

In Fig. 2 we presented the dependence of $\it h$ on $M^{2}$.
{}From this figure it follows that the best stability
region for $\it h$ is $6\,\,GeV^{2}\leq M^{2} \leq 12\,\,GeV^{2}$ ,
and thus we obtain
\begin{eqnarray}
\it f_{B^{0*}}f_{B^{0}} h&=& -0.1\, \pm 0.02   \nonumber \\
\it f_{B^{+*}}f_{B^{+}} h&=& 0.2 \, \pm 0.02
\end{eqnarray}
Note that the variation of the threshold value from $36\,GeV^{2}$
to $40\,GeV^{2}$ changes the result by few percents.
We see that the sign of the amplitudes for $B^{0}$ and $B^{+}$
are different. This is due to the fact that the main contribution
to the theoretical part of the sum rules comes from the bare loop
and the quark condensate in external field (last term in Eq. (32)).
In $B^{0}(B^{+})$ cases, both contributions have negative (positive )
signs and therefore the sign of h is negative (positive).
To get the dimensionless transition amplitude for the decay
$D^{*}\rightarrow D\gamma$, it is sufficient to make the following
replacements in Eq. (32):
\begin{eqnarray}
m_{b}\rightarrow m_{c},\,\, f_{B^{*}(B)}\rightarrow f_{D^{*}(D)},
\,\,\,\ Q_{b}\rightarrow Q_{c}, \\
and \,\,\,\,s_{0\,B}\rightarrow s_{0D}  \nonumber
\end{eqnarray}
Performing the same calculations we get the best stability region
for $\it h$ as $2\,\,GeV^{2}\leq M^{2} \leq 4\,\,GeV^{2}$ (Fig. (3)),
and we find
\begin{eqnarray}
\it f_{D^{0*}}f_{D^{0}} h&=& 0.12\,\pm 0.02   \nonumber \\
\it f_{D^{+*}}f_{D^{+}} h&=& -0.04 \,\pm 0.01
\end{eqnarray}
The signs of the transition amplitudes for $D_{0}$ and $D^{+}$
meson decays are different as in the B-meson case.

Using the transiton amplitude "h", one can calculate the decay rates for
$B^{*}(D^{*})\rightarrow B(D)\gamma$, which can be tested experimentally.
For the decay width for radiative decay $B^{*}(D^{*})\rightarrow B(D)\gamma$,
we get
\begin{eqnarray}
\Gamma(B_{0}^{*}\rightarrow B_{0}\gamma)&=& 0.28 \,KeV \\
\Gamma(B^{+\,*}\rightarrow B^{+}\gamma)&=& 1.20 \, KeV
\end{eqnarray}
and
\begin{eqnarray}
\Gamma(D_{0}^{*}\rightarrow D_{0}\gamma)&=& 14.40 \,KeV \\
\Gamma(D^{+\,*}\rightarrow D^{+}\gamma)&=& 1.50\, KeV \\
\end{eqnarray}

In order to compare the theoretical results with experimental data
for D-meson decays, we need the values of the
$D^{*}\rightarrow D\pi$ decays widths.We take these values from ref.
[13]:
\begin{eqnarray}
\Gamma(D^{*\,+}\rightarrow D^{0}\pi^{+})&=& 32 \pm 5 \,\,\,KeV \\
\Gamma(D^{*\,+}\rightarrow D^{+}\pi^{0})&=& 15 \pm 2 \,\,\,KeV \\
\Gamma(D^{*\,0}\rightarrow D_{0}\pi^{0})&=& 22 \pm 2\,\,\,KeV
\end{eqnarray}

{}From eqs.(39-43), for the BR, we obtain :
\begin{eqnarray}
BR((D_{0}^{*}\rightarrow D_{0}\gamma)  &=&  39 \% \nonumber \\
BR(D^{+\,*}\rightarrow D^{+}\gamma) &=&   3 \%
\end{eqnarray}
These results are in agreement with the CLEO data [25], which are
\begin{eqnarray}
%BR((D_{0}^{*}\rightarrow D_{0}\pi^{0})  &=&  (63.6\pm 2.3\pm 3.3) \% \nonumber
%%\\
BR((D_{0}^{*}\rightarrow D_{0}\gamma)  &=&  (36.4\pm 2.3\pm 3.3) \%  \nonumber
\\
%BR(D^{+\,*}\rightarrow D^{0}\pi^{+}) &=&   (68.1\pm 1.0) \%     \\
%BR(D^{+\,*}\rightarrow D^{+}\pi^{0}) &=&   (30.8\pm 0.4) \%     \nonumber \\
BR(D^{+\,*}\rightarrow D^{+}\gamma) &=&   (1.1\pm 1.4\pm 1.6) \%     \nonumber
 \end{eqnarray}

We see that our predictions on branching ratio are in reasonable agreement
with the experimental results.

Acknowledgement:
A part of this work was performed under TUBITAK-DOPROG Program.
One of authors
(T. M. Aliev) thanks TUBITAK for financial support.

\newpage
{ \bf Figure Captions }
\begin{description}
\item[{\bf Figure 1}:]  Diagrams contributing to the correlation function
1. Solid lines represent quarks, wave lines external currents.
\item[{\bf Figure 2}:] The dependence of the transition amplitude h
on the Borel parameter square
$M^{2}$.
Solid line corresponds $B^{0}$ and dashed line to $B^{+}$
meson cases.
\item[{\bf Figure 3}:]  The same as in Fig.2 but for
D meson case.
\end{description}

\newpage
\begin{thebibliography}{99}
\bibitem{R1} M.A.Shifman, A.I.Vainstein, V.I.Zakharov, Nucl.Phys.
B147(1979) 385.
\bibitem{R2} L.J.Reinders, H.R.Rubinstein, S.Yazaki, Phys. Rep.,
C127. (1985) 1.
\bibitem{R3} V. L. Chernyak, A. R. Zhitnitsky, Phys. Rep. 112 (1984) 173
\bibitem{R4} I. I. Balitsky, V. M. Braun, A. V. Kolesnichenko, Nucl.Phys.
B312 (1989) 509;
\bibitem{R5} V. M. Braun and I. B. Filyanov, Z. Phys. C. 48, 239 (1990)
\bibitem{R6} V. M. Belyaev, A. Khodjamirian, R. Ruckl, Z. Phys. C60
(1993) 349
\bibitem{R7} V. L. Cheryak and I. R. Zhitnitsky, Nucl. Phys. B345 (1990) 137
\bibitem{R8} P. Ball, V. M. Braun and M. G. Dosch, Phys. Rev. D44 (1991) 3567
\bibitem{R9} A.Ali, V. M. Braun and H. Simma, Z. Phys. C63 437 (1994)
\bibitem{R10} V. M. Belyaev, Z. Phys. C. 65 (1995) 93
\bibitem{R11} A.Ali, V. M. Braun, Phys. Lett. B359 (1995)
\bibitem{R12} A. Khodjamirian, G.Stoll and D. Weyler, Phys. Lett.
B358 (1995) 129
\bibitem{R13} V. M. Belyaev, V. M. Braun, Khodjamirian and R. Ruckl,
Phys. Rev. D51 (1995) 6177
\bibitem{R14} T. M. Aliev , E. Iltan and N. K. Pak, Phys. Lett. B334 (1994) 169
\bibitem{R15} T. M. Aliev , E. Iltan and N. K. Pak
(submitted for publication)
\bibitem{16}  V.A.Nesterenko and A.V.Radyushkin, Sov. J.Nucl.Phys.
39 (1984) 811
\bibitem{R17} I. I. Balitsky, V. M. Braun, Nucl. Phys. B311 (1988) 541
\bibitem{R18} V. M. Braun, I. E. Filyanov, Z. Phys. C44 (1989) 157
\bibitem{R19} V. M. Belyaev, Ya. I. Kogan, Sov. J. Nucl. Phys. (1995)
\bibitem{R20} I. I. Balitsky, A. V. Kolesnickino and A. V. Yung,
Sov. J. Nucl. Phys.  (1985)
\bibitem{R21}  T.M.Aliev, V.L.Eletsky, Sov. J. Nucl.Phys. 38 (1983) 936.
\bibitem{R22} P. Colangelo, G. Nardulli, A. A. Ovchinnikov and N. Paver,  Phys.
Lett.
B269 (1991) 204
\bibitem{R23} E. Bagan et. al., Phys. Lett. B278 (1992) 457
\bibitem{R24} M. Neubert, Phys. Rev. D45 (1992) 2451
\bibitem{R25} The CLEO Collaboration(F.Butler,et.al.),CLNS-92-
1193 (1992).
\end{thebibliography}
\end{document}